\documentstyle[prl,aps,preprint]{revtex}
\begin{document}
\title{Vortex Dynamics in Superfluids: Cyclotron Type Motion}
\author{E. Demircan$^1$, P. Ao$^2$, and Q. Niu$^1$\\
      $^1$ Department of Physics\\
      The University of Texas at Austin \\
      Austin, TX 78712 \\
      $^2$ Department of Theoretical Physics \\
      Ume\aa\ University, S-901 87, Ume\aa, SWEDEN }
\date{\today}
\maketitle
\begin{abstract}

Vortex dynamics in superfluids is investigated in the framework of the
nonlinear
Schr\"{o}dinger equation.
The natural motion of the vortex is of cyclotron type, whose frequency is found
to
be on the order of phonon velocity divided by the coherence length, and may be
heavily
damped due to phonon radiation.  Trapping foreign particles into the vortex
core can reduce
the cyclotron frequency and make the cyclotron motion underdamped.
The density fluctuations can
follow the vortex motion adiabatically within the phonon wave length at the
cyclotron frequency,
which results in a further downward renormalization of the cyclotron frequency.
We have also discussed applications on the dynamics of
vortices in superconducting films.

\ \\


\end{abstract}

\newpage
\section{Introduction}
Besides collective excitations of density fluctuations, superfluid systems can
also have topological excitations called vortices. Vortices are associated with
many important phenomena, such as quantization of circulation
\cite{vinen}, Kosterlitz-Thouless phase transition
\cite{kt}, mutual friction
\cite{DAN,donnelly}, and flux creep
\cite{donnelly,tilley}. Although the static
properties of vortices are relatively well understood, the dynamical side is
still wide open to investigations.

It has been a common practice to treat a vortex in two dimensions as a
particle, and to describe its motion by Newton's law in the following form
\begin{equation}
   M_v \ddot{{\bf r}}_0 = -2\pi \hbar \bar{\rho} \dot{{\bf r}}_0 \times
{\bf\hat{z}}
                          - \eta \dot{{\bf r}}_0,
\label{macroeq}
\end{equation}
where $ {\bf r}_0 $ denotes the vortex position, and $ \bar{\rho}$ is the
2D superfluid number density. The terms on the right hand side represent Magnus
and frictional forces respectively, and $M_v$ stands for the
vortex mass.  The Magnus force term was motivated from the behavior of vortices
in classical fluids, and has recently been related to the Berry
phase of the many-body wave function \cite{ao,gaitan,nat}.  Its existence has
been
experimentally established for superfluid $~^4$He,  but this has been a
controversial subject for magnetic fluxes in superconductors \cite{nozieres}.
The
friction term is considered to be a result of the interaction of the vortex
with collective excitations, or with the core excitations in the case of
superconductors \cite{DAN,donnelly}.
However, very little experimental information is available to test the
theories on dissipation at low enough temperatures.  The vortex mass has also
been a topic
of debate.  For incompressible classical fluids, it has been customary to
regard the vortex as
massless.  This point of view has also been adopted in some calculations for
nucleation and motion of quantized vortices \cite{volovik}.  Another point of
view is that
quantized vortices should have finite masses, roughly equal to the mass of the
vortex core \cite{muirhead}.  In references
\cite{duan,duan2,simanek,hatsuda,popov,suhl},
the vortex mass was found to be renormalized by the
condensate motion to a value logarithmically divergent with the system
size in neutral superfluids.  For charged superfluids, the vortex mass turns
out
to be finite, because the screening currents effectively replace the  system
size by the London penetration length
\cite{duan,duan2,simanek,hatsuda,popov,suhl}. Some recent
discussion on the nature of the vortex mass can be found in $\cite{nat}$. The
size of the vortex
mass may affect tunneling and specific heat of a vortex lattice
\cite{fetterexp}.

In this work, we study the vortex dynamics based on the nonlinear
Schr\"{o}dinger equation, which has been used to model
superfluids in a semi-microscopic manner. This equation has been derived by
Gross and Pitaevskii for a weakly interacting superfluid \cite{gross,ll}.
In the appendix, we  derive a nonlocal version of the
nonlinear Schr\"odinger theory  from Feynman's
many-body trial wave function, which can take into account strong correlations
in
a superfluid such as  $~^4$He.  The derivation of the nonlinear
Schr\"{o}dinger equation for superconductors in the clean limit is given in
Refs.\cite{ao2,aitchison}.  This equation contains both solutions for
collective excitations of
density fluctuations and for topological excitations of vortices.
It provides a useful starting point
for the study of vortex motion and its coupling to the collective excitations.
An effective  Lagrangian will be derived for
the  vortex coordinate and for the density and phase fields of the superfluid
condensate.  As we will see, the equation of motion for a vortex naturally
contains the
Magnus force, and the effect of the condensate motion.

We will concentrate our attention on the cyclotron motion of the vortex.
Such a motion is a natural solution of the phenomenological equation (1), with
the cyclotron
frequency given by
\begin{equation}
    \omega = \frac{2\pi \hbar \bar{\rho}}{M_v} \label{omegamass}
\end{equation}
if damping is ignored.  We found that the cyclotron motion is also a natural
solution of the equations of motion of the vortex based on the nonlinear
Schr\"{o}dinger equation.
We found that the cyclotron motion is damped by phonon radiation from the
vortex.  For a bare vortex, the damping rate is found to be about
the same as the cyclotron frequency.  For a vortex with trapped particles in
it,
the damping can be much smaller than the cyclotron frequency, although the
cyclotron frequency itself is also reduced.
 With the cyclotron frequency determined from the equations of motion for the
vortex coupled to the condensate, we can use equation (2) to define the vortex
mass. The frictional coefficient $\eta$ in equation (1) can be calculated from
the
rate of damping due to phonon radiation.

We have carefully examined the adiabaticity assumption used in references
\cite{duan,duan2,simanek,popov,suhl} that the field of condensate phase
follows rigidly with the moving vortex.  We found that this is possible
within a length scale of $\lambda \approx \xi (1+ M_e/M_c)$, where $\xi $ is
the coherence
length, and $M_c $ and $M_e $ are the masses of the vortex core and of the
trapped particles.
Outside this length scale, the condensate cannot follow the vortex motion.
Because of this, we
found that the logarithmic divergence of the vortex mass with the system size,
which was found in
references \cite{duan,duan2,simanek,popov,suhl}, is cut off by the scale of
phonon wavelength,.

The organization of the paper is as follows. In Sec. II, we will give the basic
ingredients of the nonlinear Schr\"odinger theory, and obtain the static vortex
solution. In
Sec. III, we will introduce vortex motion, derive the effective Lagrangian and
the
dynamical equations for the vortex and condensate. In Sec. IV, we will solve
for the condensate
response to the cyclotron motion, and find the conditions when the adiabaticity
assumption is valid.
In Sec. V, we will study the low frequency motion of the vortex, which is
possible
when the mass of the trapped particles is large. In Sec. VI, we present our
numerical results
for finite frequency of the cyclotron motion.  In
Sec. VII we will present our conclusions.


\section{Nonlinear Schr\"{o}dinger Lagrangian}

Our starting point is the nonlinear Schr\"{o}dinger equation which may be
derived
from the following Lagrangian:
\begin{equation}
    L = \int d^2 r \left[  i \hbar \psi ^{\ast} \frac{\partial}{\partial t}
\psi
          - \frac{\hbar ^2}{2 m} \left| {\nabla} \psi \right| ^2
          - \frac{1}{2} V \left[ \left| \psi \right | ^2 -
                                  \bar \rho
                        \right]^2 \right]  , \label{gp}
\end{equation}
where $ m $ stands for the mass of the superfluid atom, $ V $ represents the
interaction potential between the atoms, and $
\bar{\rho} $ is the background superfluid number density. The sign of $ V $ is
positive to represent a repulsive interaction.

The natural length and time scales of the nonlinear Schr\"odinger equation are
$ \xi=\hbar/(m V \bar{\rho})^{1/2}$ and  $ \tau =\hbar/(V\bar{\rho}) $.  As
will be seen later,
$\xi$ gives the length scale of a vortex, and $\tau$ gives the time scale of a
bare vortex in
cyclotron motion. If we scale the Lagrangian by the energy $ \hbar ^2
\bar{\rho}/ m$, then the
Madelung transformation $ \psi = \sqrt{\rho} e^{iS} $ puts the Lagrangian in
the dimensionless form
\begin{equation}
      L = -\int \left[  \rho \dot{S}
          + \frac{1}{2} \rho \left| {\nabla} S \right|^2
          + \frac{1}{8\rho} \left| {\nabla} \rho \right|^2
          + \frac{1}{2}     \left[ \rho - 1 \right]^2 \right] d^2 r .
\label{gpl}
\end{equation}
The dynamical equations of the condensate follow from the variation of the
action with respect to the phase and the density
\begin{equation}
   \dot{\rho} + {\nabla}\cdot [ \rho {\nabla} S ]  =  0,
\end{equation}
\begin{equation}
   \dot{S} + \frac{1}{2} |{\nabla} S |^2 + \frac{|{\nabla}
\rho |^2}{8 \rho ^2}  - \frac{\nabla ^2 \rho}{4 \rho} + \rho - 1   =  0.
\end{equation}
The first equation is nothing but the continuity equation, whereas the second
resembles the Euler equation of hydrodynamics of classical fluids.
The normal modes of the  linearized equations of the fluctuations around the
uniform condensate describe the collective excitations of the system.
The low frequency part of the spectrum is phonon like with velocity $c_s =
\xi/\tau =\sqrt{V\bar{\rho}/m}$ \cite{DAN,ll}. However, the roton minimum is
absent in the spectrum,
because the short range atomic repulsion is not appropriately taken into
account. In the Appendix,
we give the derivation of an effective Lagrangian that may overcome this
problem.

The equations also allow vortex solutions.  For simplicity, we consider a
single vortex with unit circulation around the center $
{\bf{r}}_0$ in the $x-y$ plane, with the phase $ S $ of the condensate
wave function given by
\begin{equation}
    S = S _0 \equiv \Theta({\bf{r}} - {\bf{r}}_0) \label{s0},
\end{equation}
where $ \Theta = \mbox{arctan}[(y-y_0)/(x-x_0)] $ is the polar angle of $
{\bf r} - {\bf{r}}_0$.  The density $ \rho_0 $ satisfies
\begin{equation}
   \frac{|{\nabla}\rho_0 |^2}{8\rho_0^2}
 - \frac{\nabla ^2 \rho_0}{4 \rho_0} + \rho _0 - 1
 + \frac{1}{2 |{\bf{r}}-{\bf{r}}_0|^2}   = 0.
\label{rho0}
\end{equation}
The asymptotic forms are easy to find:
\begin{equation}
   \rho_0 (r) =
  \left\{
    \begin{array}{lll}
         2r^2 & , & r \ll 1 ; \\
         1 -\frac{1}{2r^2}& , & r \gg 1, \label{rhoapp}
    \end{array}
  \right.
\end{equation}
A simple analytical expression that interpolate between the above asymptotic
forms
is given by \cite{clem}
\begin{equation}
   \rho_0 (r) = \frac{2r^2}{1 + 2r^2}.\label{rho0a}
\end{equation}

\section{Moving Vortex}

In this section we will derive the equations of motion for a moving vortex.  We
assume that the motion of the vortex has a small amplitude (not necessarily
slow), such
that the fields of density and phase of the condensate may be expanded as
\begin{eqnarray}
       S & = & S_0    ( {\bf{r}} - {\bf{r}}_0 (t) ) + S _1({\bf {r}},t)
,\nonumber
\\
    \rho & = & \rho _0( {\bf{r}} - {\bf r}_0 (t) ) + \rho _1({\bf r},t) ,
\label{spd}
\end{eqnarray}
where at any instant of time, $ S_0 $ and $ \rho_0 $
satisfy the static vortex equations (\ref{s0}) and (\ref{rho0}). $ S _1 $
and $ \rho _1 $ represent small corrections caused by the motion of the vortex.
By substituting Eq.(\ref{spd}) into the Lagrangian (\ref{gpl}), we arrive at
an effective Lagrangian for $S _1({\bf {r}},t)$ and $\rho _1({\bf r},t)$ as
well
as the vortex coordinates ${\bf{r}}_0 (t)$.  There is no problem of redundancy
in
the dynamic variables; unlike the original phase field $S({\bf {r}},t)$,
$S_1({\bf {r}},t)$ is required to be single valued \cite{zhang}.
Keeping up to second order
terms in ${\bf r}_0$, $S_1$ and $\rho_1$, we find the new Lagrangian
as
\begin{eqnarray}
    L  & = & -\int  d^2 r \left[
               \rho_{0} \dot{S}_0 + \frac{1}{2} \rho_0 | {\nabla} S _0 |^2
             + \frac{1}{2} [ \rho _0 - 1 ]^2  \right.
             + \rho_0  \dot{S} _1 +  \rho _1 \dot{S}_0 + \rho_0
{\nabla} S_0 \cdot {\nabla} S _1  + \nonumber \\
       &   & + \frac{1}{2}|\nabla S_0 |^2 \rho_1 + [\rho_0 - 1 ]\rho_1
             + \rho _1 \dot{ S} _1 + \frac{1}{2} \rho_0 |{\nabla}
 S _1 |^2 + \rho _1 {\nabla} S_0 \cdot {\nabla} S _1
             + \frac{1}{2} \rho_1 ^{2}
\nonumber \\
       &   & + \left. \frac{1}{8 \rho _0}|{\nabla} \rho_0|^2
             + \frac{1}{4 \rho _0}{\nabla} \rho_0 \cdot {\nabla}\rho _1
             - \frac{|{\nabla} \rho_0|^2}{8\rho _0 ^2} \rho _1
             + \frac{|{\nabla} \rho_0|^2}{8\rho _0 ^3}{\rho _1} ^2
             - \frac{\rho_1}{4\rho _0 ^2}{\nabla} \rho_0 \cdot
{\nabla}\rho _1
             + \frac{| {\nabla} \rho _1|^2 }{8\rho_0}
       \right] .
\end{eqnarray}

We will add an extra kinetic energy term of the form
\begin{equation}
        \frac{1}{2} M_e \dot{{\bf r}}_0 ^2,
\end{equation}
to simulate the situation with particles trapped inside the vortex core.
It is understood that this may be an over simplification of
the  real interactions between the trapped particles and the vortex
\cite{trap}, but the
main purpose of introducing this term is to provide a mechanism for controlling
the time scale of  the vortex motion.  The external mass $ M_e$ is measured in
units of $(m\xi^2 \bar{\rho})$, and a bare vortex is described by $M_e = 0 $.

The resulting Lagrangian can be simplified considerably by using the
equations for the static vortex, integration by parts and dropping the
constant terms:
\begin{eqnarray}
    L  & = & -\int  d^2 r \left[
                     \rho_{0} \dot{S}_0 + \rho_0  \dot{S }_1
                + \rho _1 \dot{S}_0 + \rho _1 \dot{ S }_1 \right.
                + \frac{1}{2} \rho_0 |{\nabla} S _1 |^2
                + \rho _1 {\nabla} S_0 \cdot {\nabla} S _1 +
\nonumber \\
       &   &
                + \frac{|{\nabla} \rho_0|^2}{8 \rho_0 ^3}{\rho _1}^2
                - \frac{\rho _1}{4\rho _0 ^2} {\nabla} \rho_0 \cdot
{\nabla} \rho_1
                + \frac{1}{8 \rho_0}|{\nabla} \rho_1|^2
                + \frac{1}{2}\left. {\rho _1}^2
                             \right] - \nonumber \\
       &   &    - \oint S_1 \rho_0 \nabla S_0 \cdot {\bf \hat{n}} d\ell +
\frac{1}{2}M_e
\dot{{\bf r}}_0^2  .
\end{eqnarray}
The boundary conditions will be taken as $\rho_1 = 0 $ and $\nabla S_1 =0$ as
well
as $\dot{S}_1 = 0$ at infinity. Then, the line integral, which is taken around
the boundary,
only adds a constant to the Lagrangian and can be dropped. The dynamical
equation of the vortex is
then obtained by variation of the action with respect to $ {\bf r}_0$. The
equation,
linearized in $\dot {\bf r}_0$, $ \rho_1 $ and $ S_1$, is
\begin{equation}
    - 2\pi \dot{{\bf r}}_0 \times {\bf \hat{z}}  +
      \int \left[  \dot{S} _1 {\nabla} \rho _0
                 - \dot{\rho} _1 {\nabla} S_0
           \right] d^2 r    - M_e \ddot{{\bf r}}_0 = 0,
\label{voreq}
\end{equation}
where we have used the fact that ${\nabla} _{{\bf r}_0}$ can be replaced
by $ -{\nabla} $ when it acts on $( {\bf r} - {\bf r}_0 )$. The first term
represents the well-known Magnus force. The second term shows the coupling
between the vortex and the condensate, and the last term is the usual inertial
force on the particles trapped at the core. Note that by comparing
Eq.(\ref{voreq}) to Eq.(\ref{macroeq}), one can extract the
phenomenological parameters, such as the vortex mass and coefficient of
viscosity.  Also note that, it is not possible to identify the
vortex mass immediately from this equation without knowing the perturbations $
\rho_1 $ and $ S_1 $.

The linearized equations for $ \rho _1$ and $ S_1 $ are
\begin{equation}
   \dot{S}_1 + {\nabla}S_0 \cdot {\nabla}S _1
             + \frac{1}{4\rho_0 ^2} {\nabla}\rho_0 \cdot
                    {\nabla}\rho_1
             - \frac{1}{4\rho_0} \nabla ^2\rho _1
             - \frac{|{\nabla}\rho_0|^2}{4 \rho _0^3}\rho _1
             + \frac{\nabla^2 \rho _0}{4 \rho_0 ^2}\rho _1
             + \rho _1
             =  \dot{{\bf r}}_0 \cdot {\nabla}S_0,
\label{coneq1}
\end{equation}
\begin{equation}
\dot{\rho}_1 + {\nabla}\rho _0 \cdot {\nabla}S_1 + \rho_0 \nabla^2 S_1
             + {\nabla} S_0 \cdot {\nabla}\rho _1
             =  \dot{{\bf r}}_0 \cdot {\nabla}\rho _0.
\label{coneq2}
\end{equation}
{}From these two equations, we notice that the dynamics of the condensate is
driven by the motion of the vortex.  In other words, the vortex is accompanied
with
backflow-like  corrections, $\rho_1$ and $S_1$, which on the other hand, act
back
on the vortex in a way determined via Eq.(\ref{voreq}).

In the rest of the paper, we choose the origin of ${\bf r}$ to be
instantaneously at $ {\bf r}_0 $. The dynamical equations (\ref{voreq}),
(\ref{coneq1}) and (\ref{coneq2}) remain unchanged up to second order terms in
$
{\bf r}_0$, $ \rho_1 $ or $ S_1 $.  Also, the boundary conditions are
such that $\rho_1$ and $S_1$ are regular at the origin, while $ \rho _1
\rightarrow 0 $, $ \nabla S_1 \rightarrow 0 $ at infinity (boundary).

\section{The Equations for Cyclotron Motion}

We would like to consider solutions of the following form:
\begin{equation}
   {\bf r}_0  =  \hbox{Re}\left[ b e^{-i \omega t}({\bf \hat{x}} + i
{\bf \hat{y}})
\right] ,
\label{r0}
\end{equation}
\begin{equation}
     \rho _1 = \mbox{Re}\left[  b F(r) e^{-i(\omega  t - \theta)} \right]
\label{rho},
\end{equation}
\begin{equation}
        S _1 = \mbox{Re}\left[ ib G(r) e^{-i(\omega  t - \theta)} \right]
\label{S}.
\end{equation}
 The constant parameters $ \omega  $ and $ b $, and the functions $
F(r) $ and $ G(r) $, are yet to be determined.  The first expression
shows that the motion of the vortex is of cyclotron type with size $b$ and
frequency $\omega$.  The frequency is allowed to have an imaginary part to
describe a damped cyclotron motion.  A complex phase factor in $b$ is
immaterial,
because it only affects the initial angle of ${\bf r}_0$.  A simple angular
harmonic analysis shows that the angular dependence of the condensate response
has to be in the given form.
In a general sense, this motion is the analogue of the massive branch of the
helical vortex waves in classical fluids \cite{lamb,fetter}.
For simplicity the expression ``Re"
will be dropped in the rest of the discussion.

Using the following expressions for the right hand sides of
equations (\ref{coneq1}) and (\ref{coneq2})
\begin{eqnarray}
       \dot{{\bf r}}_0 \cdot \frac{{\bf \hat{\mbox{\boldmath $\theta$}}}}{r}
                 & = & \frac{b\omega}{r} e^{-i(\omega  t - \theta)},
\nonumber \\
   \dot{{\bf r}}_0 \cdot {\bf \hat{r}} \rho_0 '
                 & = & -i b\omega  \rho_0 ' e^{-i(\omega  t - \theta)},
\end{eqnarray}
we find that Eqs.(\ref{voreq}),(\ref{coneq1}) and
(\ref{coneq2}) imply the following equations for $ \omega $, $ F $, and $ G $:
\begin{equation}
    \frac{1}{2} \int_0 ^{\infty} \left[ r \rho_0' G + F \right] dr
      + \frac{M_e}{2\pi} \omega  =  1 ,
\label{vorcon}
\end{equation}
\begin{equation}
  \frac{\rho _0 '}{4 \rho_0 ^2} F'
    - \frac{1}{4\rho_0}\left[ F'' + \frac{F'}{r} -\frac{F}{r^2}\right]
    - \left[ \frac{{\rho_0 '}^2}{4 \rho_0 ^3}
           - \frac{\rho_0 '' + \rho_0 ' /r}{4 \rho_0 ^2} -1
      \right] F + \left[ \omega - \frac{1}{r^2}
                  \right]G
            = \frac{\omega}{r} ,
\label{FG1}
\end{equation}
\begin{equation}
   \rho_0 ' G'
   + \rho_0 \left[ G'' + \frac{G'}{r} - \frac{G}{r^2}
            \right] - \left[ \omega - \frac{1}{r^2} \right] F
            =  - \omega \rho_0 ' ,
\label{FG2}
\end{equation}
where a prime denotes differentiation with respect to $ r $.
Note that the parameter $ b$ has been factored out from the above equations,
and it is only to be fixed by the initial velocity of the vortex.

\section{Condensate Response to the Cyclotron Motion}

In this section, we consider the response of the condensate density and phase
to the vortex motion.
Analytic and semiquantitative results may be obtained  by studying the limits
of large ($ r\gg 1$)
and small distances ($ r\ll 1$). The large $ r $ limit carries the information
on how the system size may affect the dynamics, and the small $ r $ limit gives
the contribution
of the core where most of the variation in $ \rho _0 $ occurs.   These two
limits cover the essential features of the dynamics.

When $ r \ll 1$, we can neglect $ \omega $ and
replace $\rho_0$ by $2r^2$ (cf.Eq.(\ref{rhoapp})) on
the  left hand side of Eqs.(\ref{FG1}) and (\ref{FG2}). After a straightforward
but
tedious calculation one can show that the solutions are of the following form
\begin{equation}
    F(r) = 4 A \omega  r^3, \label{fsmall}
\end{equation}
\begin{equation}
    G(r) = B \omega r . \label{gsmall}
\end{equation}
It is not difficult to find $ A + B = -1 $,  although the individual values of
$ A$
and $ B $ are undetermined yet.  Our knowledge of their sum will be enough for
the purpose
of evaluating the integrand in Eq.(\ref{vorcon}).

When $ r \gg 1$, we may use the approximation of $\rho_0 $ for large $r$ as
given in
Eq.(\ref{rhoapp}).  Also, the first three terms on the left hand side of
Eq.(\ref{FG1}) is dominated
by $F$, yielding
\begin{equation}
 F + \left[\omega-{1\over r^2}\right] G
   =  \frac{\omega}{r}
\label{FGr1} .
\end{equation}
Substituting this relation into Eq.(\ref{FG2}), we arrive at the following
equation for $ G$
\begin{equation}
   G'' + \frac{1}{r} G' - \frac{1+2\omega}{r^2} G + \omega^2 G =
         \frac{\omega ^2}{r}-{2\omega\over r^3}.
\label{FGr2}
\end{equation}
It is easily verified that a special solution is $G = \frac{1}{r}$.
To obtain the general solution, we note that the corresponding homogeneous
equation for $G$ is
that of the Bessel functions of order $\nu=\sqrt{1+2\omega}$.  We only keep the
outgoing
wave, which represents the radiated phonons, then the general solution is given
by
\begin{equation}
   G = \frac{1}{r}+CH_\nu^{(1)}(\omega r), \label{glarge}
\end{equation}
where $C$ is a constant and $H_\nu^{(1)}$ is the first kind Hankel's function.
This implies via Eq.(\ref{FGr1})
\begin{equation}
   F = \frac{1}{r^3}-C\left[\omega -\frac{1}{r^2}\right]H_\nu^{(1)}(\omega r),
\label{flarge}
\end{equation}
A similar radiation process is also found for an oscillating object in a
classical fluid
\cite{fetterbook}.

The constants $B$ and $C$ must be determined by matching the solutions in the
intermediate
region $r\approx 1$.  For semiquantitative purposes, we may regard the above
solutions for small
and large $r$ to be valid up to the coherent length $r=1$ (where $\rho_0=2/3$).
Then the continuity of $G$ and its derivative yield
\begin{equation}
C={ 2\over  \omega {H_\nu^{(1)}}'(\omega)-H_\nu^{(1)}(\omega)}
\end{equation}
and
\begin{equation}
B={1\over \omega } {\omega {H_\nu^{(1)}}'(\omega)+H_\nu^{(1)}(\omega)
\over \omega {H_\nu^{(1)}}'(\omega)-H_\nu^{(1)}(\omega)},
\end{equation}
where a prime denotes differentiation with respect to the argument.

\section{Low Frequency Solutions}

It will be instructive to consider the case of small $\omega$, which, as we
will see later,
may be achieved in the limit of large mass of the trapped particles.  Using the
expansion of
the Hankel function for the limit of  small argument, we can find the
constants as $C=-i{\pi\over 2}\omega$, and $B={1 \over 2}$.

Depending on the assumption that the asymptotic forms of the solutions are
valid in the whole
regions above and below $r=1$ respectively, the cyclotron frequency can be
determined from Eq.(\ref{vorcon}). Obviously, this constitutes a rough estimate
of the frequency,
but we expect the qualitative behavior to be well reflected. Then, to lowest
order in $\omega$ we
have
\begin{equation}
     \left[{1\over 2} \ln(1/ \omega)+{\pi\over 4}i\right] \omega
 + \frac{M_e}{2\pi} \omega =1.
\label{omegaeq}
\end{equation}
In the large $M_e$ limit, we can solve for $\omega$ as
\begin{equation}
   \omega =  {1\over {M_e\over 2\pi}+{1\over 2}\ln({M_e\over 2\pi})+{\pi\over
4}i}
\label{omegasimp}.
\end{equation}
This gives a vortex mass via Eq.(2)
\begin{equation}
   M_v = {M_e} + \ln({M_e\over 2\pi}) + {\pi ^2 \over 2} i.
\end{equation}
Apart from the mass of the trapped particles, there is a hydrodynamic
correction (the second
term) which diverges logarithmically with $M_e$.   The imaginary part
represents a decay rate of
the cyclotron motion due to  the  radiation of phonons. It is seen that this
decay becomes
unimportant when  $M_e>>\pi^2/2$.

To understand the logarithmic correction, we divide the $r>1$
region further into two regimes:  $1<r<1/\omega$, and $r> 1/ \omega$.
In the former regime, we can use the
asymptotic form of the Hankel's function, with the result:

\begin{equation}
   G = {0.575 \omega\over r},
\end{equation}
and
\begin{equation}
   F = {\omega\over r}.
\end{equation}
This gives a density response $\rho_1$ exactly the same as found using the
adiabaticity assumption
\cite{duan,duan2,simanek,popov}. We will therefore call this regime as the
adiabatic regime.  This
regime has a size of one phonon wavelength $\lambda={1\over \omega}$, and can
be very large when
$M_e$ is large. The adiabatic following of the condensate with the vortex
in this region is responsible for the logarithmic correction to the vortex
mass.

For $r >> 1/\omega $ we can write the solutions as
\begin{equation}
   G = {1\over r}-i\sqrt{\pi\omega\over 2r}e^{i(\omega r+{\pi\over 4})} ,
\end{equation}
and
\begin{equation}
   F = {1\over r^3}+i\omega \sqrt{\pi\omega\over 2r}e^{i(\omega r+{\pi\over
4})} .
\end{equation}
It is not difficult to show that to first order in ${\bf r_0}$, the condensate
phase and density
becomes
\begin{equation}
    S({\bf r},t) = S_0({\bf r})
          + b\sqrt{\pi\omega\over 2r}e^{i(\omega r+{\pi\over 4})} e^{-i\omega
t},
\end{equation}
and
\begin{equation}
    \rho({\bf r},t) = \rho_0({\bf r})
          + i b\omega \sqrt{\pi\omega\over 2r}e^{i(\omega r+{\pi\over 4})}
e^{-i\omega t}.
\end{equation}
The first terms on the right hand sides are the phase and density of a static
vortex at the
origin, and the other terms are oscillations in the condensate that represent
the phonons radiated
by the motion of the vortex. Therefore, at distances larger than $ \lambda $
the motion of the
condensate is not an adiabatic following of the vortex.

We now calculate the coefficient of viscosity corresponding to the damping. It
is given by
\begin{equation}
   \eta = \frac{2\pi \hbar \bar{\rho} }{Q},
\end{equation}
where $Q$ is the quality factor, and we have restored the real units. Using the
previous
results, we find $Q=M_c/M_e$, where $M_c={\pi^2\over 2}\xi^2 m\bar \rho$ is
roughly
the mass of the superfluid that is expelled from the vortex core.  Using the
value
of the three
dimensional density $\bar{\rho}_{3D}  \approx 10^{28} \mbox{m}^{-3} $ for
superfluid He$^4$, the
viscosity per unit length of a vortex line becomes
\begin{equation}
   \frac{\eta}{d} \approx 7 \times 10^{-6}  \left[ \frac{M_c}{M_e} \right]
\mbox{kg}/(\mbox{m} \cdot \mbox{s}).
\end{equation}
In comparison to the viscosity induced by the scattering of excitations in
superfluid $~^4$He at temperature 1 K \cite{donnelly}, the numerical factor is
on the same
order, but the factor $(M_c/M_e)$ can reduce it further.


\section{Cyclotron Motion of Finite Frequency}

In order to extend our conclusions to the small $M_e$ as well as to the bare
vortex case, we
adopt a semi-numerical method. We use the approximate solutions found in
Eqs.(\ref{fsmall}),(\ref{gsmall}),(\ref{glarge}) and (\ref{flarge}) to find a
numerical solution
of the cyclotron frequency from Eq.(\ref{vorcon}). Our results are contained in
Fig.(\ref{omegafig}), where solid lines are the real and imaginary parts of the
cyclotron
frequency from the numerical calculation, and the dashed lines are from the
approximate expression
in Eq.(\ref{omegaeq}).

The agreement between the approximate and numerical solution for large $M_e$ is
quite
satisfactory. The estimate, $M_e >> \pi^2 /2 $, for the radiative damping to be
negligible is
seen to be appropriate. In general we see that as $M_e $ decreases, the
magnitudes of the imaginary and real parts increase, and for $M_e =0 $ they are
both
approximately given by the time scale of the nonlinear Schr\"odinger Lagrangian
\begin{equation}
 \omega_{r,i} \approx \tau^{-1} = \frac{\hbar}{m\xi ^2}.
\end{equation}
Thus, the motion of a bare vortex is heavily damped. Also, we notice that
the imaginary part is smooth over the whole range, but there is a peak in the
real part near
$M_e /2\pi \approx 0.81$. This can either be a superficial defect caused by the
inappropriate
handling of the solutions at $r\approx \xi $, or a real physical phenomenon,
the nature of which
can only be substantiated by a numerical solution of the complete dynamical
equations. Such an
effort is currently underway.


\section{Discussion and Conclusions}

In summary, we have tried to give a better understanding of the dynamics of
vortices in superfluid systems. We have derived an effective Lagrangian
from a nonlinear Schr\"{o}dinger Lagrangian, and obtained the dynamical
equations
for the vortex coupled with the condensate phase and density.

We showed that the natural motion of the vortex turns out to be of
cyclotron type just like the one predicted by the phenomenological
vortex equation in (\ref{macroeq}).  There are three qualitatively different
regimes for the condensate response: the core regime ($r<\xi$), the adiabatic
regime ($\xi<r<\lambda$), where $\lambda$ is the phonon wavelength at the
cyclotron frequency,
and the radiation regime ($r>\lambda$). Combined with the numerical result,
this wavelength is
roughly given by $ \lambda \approx \xi (1+ M_e /M_c)$. When the mass of the
trapped particles is
large compared to the core mass, the cyclotron frequency is low, and there is a
 logarithmic
correction to the vortex mass due to the adiabatic following of the  density
and phase
fluctuations in the large region of adiabatic regime \cite {duan}. Similar
results have been
obtained by Wexler and Thouless \cite {Aspen}, and Arovas and Freire
\cite{arovas} using different
methods.  The phonon radiation damping is negligible here as was expected in an
earlier work
of Niu, Ao and Thouless \cite {nat}.

For a bare vortex, the cyclotron period is on the order of the time that a
phonon
travels a coherence length, and the vortex mass is on the order of the mass of
the
fluid that can occupy the core.  The adiabatic regime is essentially empty, and
the
logarithmic correction is absent.  However, due to the large density of states
of the
phonons at the enhanced frequency, radiation damping is heavy and may
completely
overshadow the cyclotron motion.

In order to experimentally observe the vortex cyclotron motion, one has to
create a situation
of small damping.  The present theory is too crude to exactly tell whether a
bare vortex in
superfluid  $~^4$He should be overdamped or underdamped, because the nonlinear
Schr\"odinger
equation does not treat the core structure accurately.  One needs to carry out
a microscopic
calculation in order to pin down this issue.   In the appendix, we briefly
outline one such
method.

Ion trapping in a vortex core can be a useful way of reducing the cyclotron
frequency
and the damping.  We emulated this effect crudely by adding an external mass
term to the
Lagrangian.  In reality, a trapped ion also expel the superfluid particles,
making the core
much bigger than the coherence length $\xi$.  For example, for negative ions
(electrons)
the expanded core size can be as large as $16 \AA$ compared to a bare core size
of $1 \AA$ in
superfluid $~^4$He.  In such cases the external mass in our theory should also
include the
expelled  superfluid mass by the ions.  The vortex can be driven either by an
oscillating
superflow, or by an ac electric field which acts on the ions.  Resonances
should
be observed at the cyclotron frequency, which is on the order of $\hbar \over m
a^2$ where
$a$ is the radius of the hollow core.  For the case of a negative ion, this
gives
$\omega \approx 6$ GHz.

When we consider charged superfluids, we must deal with the complication due to
the existence of yet another length scale, the London penetration length.
However, for thin
superconductor films, the London penetration length is quite large, and
vortices are very
similar to their counterparts in $^4$He.  For a bare vortex, the adiabatic
length is
about the same as the coherence length, so that there is no logarithmic
correction due to the
adiabatic following of the condensate.  The cyclotron frequency should be given
by the
flux quantum times the background 2D charge density divided by the vortex mass:
\begin{equation}
    \omega={\hbar\over m_e\xi^2},
\end{equation}
which only depends on the electron mass and the coherence length.
Using the relation between the coherence length and the superconducting gap
$\Delta$, $\xi={\hbar
v_f\over \pi \Delta}$, the cyclotron energy is found to be smaller than the gap
by a small factor
equal to $\Delta/\epsilon_f$, where $\epsilon_f$ is the normal-electron Fermi
energy.
The cyclotron motion may be resonantly excited by an ac superflow as we
discussed above for the
case of  superfluid $~^4$He.  For Al thin films, in the very clean limit,
$\xi=1.6$ $\mu$m, we
have $\omega=45$ MHz.   The resonance should show itself as a peak in the
resistance-frequency
plot.

The finite temperature effects can also be foreseen qualitatively along the
lines of the present
analysis. Primarily, since the core size increases with the temperature, the
vortex mass should
become larger, and one expects to see the resonances at lower cyclotron
frequencies. Another
fact that contribute to this effect is the reduction of the Magnus force due to
the reduction of
the superfluid density at finite temperature.

Similar to the mechanism of phonon radiation damping in neutral superfluids,
there is plasmon
radiation damping in the  superconductor case, but the effect is not strong
because of low (zero
for 3D) density of states of plasma at the cyclotron frequency.  The cyclotron
motion may also
be damped by the excitations of normal electrons in the core.  This effect
might be strong due
to the coincidence of the cyclotron energy with the excitation energies of
normal electrons
in the core, but the exact nature of this coupling has to be established in a
microscopic
calculation.

\section{Acknowledgments}
Part of this work was done at the Aspen Center for Physics in the summer of
1995, where
we received many instructive comments from David Thouless and Sandy Fetter.  We
are also indebted
to Michael Stone, Daniel Arovas for many stimulating discussions, and to G. A.
Georgakis, M. C.
Chang, and A. Barr for valuable comments. E. D. is also grateful for
discussions with J. A.
Freire. This work is supported by the Welch Foundation.

\section{Appendix: A nonlocal nonlinear Schr\"odinger Theory}

Assume that we are given the correct ground state $\left|0 \right >$ of a
superfluid in the
absence of vortices and collective excitations.  Then to a good approximation,
the dynamics
of the superfluid may be described by the Feynman wave function \cite{feynman}:
\begin{equation}
     \left| \Psi _{F} \right> = \frac{\exp {(  \sum_j( i S({\bf r}_j) +
                                      \alpha({\bf r}_j)) )}
                                      \left| 0 \right >}
         { \left[ \left< 0 \right|\exp({ 2\sum _j \alpha({\bf r}_j) })
                  \left| 0 \right >
           \right]^{1/2} } ,
\end{equation}
where $ S $ and $ \alpha $ are time-dependent real functions,
and the denominator is a normalization factor.
The essential feature of the Feynman wave function lies in the
factor of products of single particle wavefunctions, whose position
and time dependence may generate superflow, density fluctuations, and
quantized vortices.

The dynamical equations for $S$ and $\alpha$ may be obtained variationally.
The time-dependent Schr\"odinger equation can be
regarded as the Euler-Lagrangian equations of the following Lagrangian
\begin{equation}
      L =  i \left< \Psi \right|
               \frac{\partial}{\partial t}
           \left| \Psi \right>
          -\left< \Psi \right| H \left| \Psi \right>.
\end{equation}
where $H$ is the full microscopic Hamiltonian containing the kinetic as well as
interaction energies of the superfluid particles.  If we substitute the
Feynman wave function into the above expression, then, after some mathematical
manipulations,
the following  Lagrangian may be obtained for the fields $S$ and $\alpha$:
\begin{equation}
     L = -\int \rho({\bf r})
         \left[ \dot{S}({\bf r})
         + \frac{1}{2} \left| \nabla S({\bf r}) \right| ^2
         + \frac{1}{2} \left| \nabla \alpha ({\bf r})\right| ^2
         \right] d^2r,
\end{equation}
where the density $ \rho({\bf r}) $ is defined by
\begin{equation}
   \rho ({\bf r}) = \frac{ \left< 0 \right|
                 \exp{(2 \int \alpha({\bf r}') \hat{\rho}({\bf r}') d^2 r')}
                 \hat{\rho}({\bf r})\left| 0\right>
                         }
                 { \left< 0 \right|
                 \exp{(2 \int \alpha({\bf r}) \hat{\rho}({\bf r})d^2 r)}
                   \left| 0 \right >
                 } \label{FL}
\end{equation}
with $\hat \rho$ being the density operator.
The last two equations provide a self-contained set. Incorporated with the
works on the ground state wavefunctions \cite{microgs}, this procedure
constitutes a microscopic
derivation of an effective Lagrangian that takes care of the short distance
effects in a better way
than the nonlinear Schr\"odinger Lagrangian.

The usual nonlinear Schr\"odinger Lagrangian is recovered if we take $\left| 0
\right >$
as the simple product state of zero momentum.  In this case, the density
$\rho({\bf r})$
is simply $\exp(2\alpha({\bf r}))$.  In the general case, Eq.(\ref{FL}) still
resembles
the usual nonlinear Schr\"odinger Lagrangian by having identical terms
involving the phase
field $S$.  The term involving $\alpha$ is generally a nonlocal and nonlinear
functional
of $\rho$, but is independent of $S$.  A full scale study of the implementation
and
implications of the nonlocal nonlinear Schr\"odinger theory will be presented
in separate
publications.



\begin{figure}
\caption
{Plot of the real (upper curves) and imaginary parts (lower curves) of the
cyclotron frequency
as a function of the external mass, calculated by using the analytical
solutions (solid) and by
using Eq.(33) (dashed). The frequency $\omega $ is in units of $\tau^{-1}$, and
the external mass
$ M_e / 2\pi $ is in units of $ \xi^2 \bar{\rho} m $. }
\label{omegafig}
\end{figure}

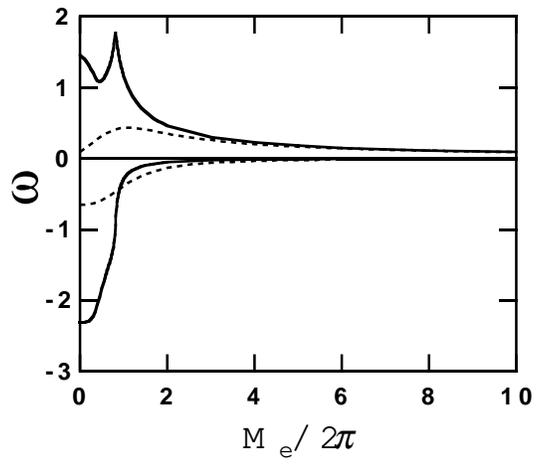
\end{document}